\documentclass[prl,aps,twocolumn,superscriptaddress,showpacs,preprintnumbers,amsmath,amssymb]{revtex4}
\usepackage{color}
\usepackage{graphicx}
\usepackage{ulem}
\usepackage{float}
\usepackage{here}
\usepackage{url}
\usepackage[colorlinks=true,urlcolor={blue},citecolor={blue}]{hyperref}
\usepackage{dcolumn} 
\usepackage{bm, bbm}      
\usepackage{curves}
\usepackage{epic}
\usepackage{wasysym}
\usepackage{subfigure}
\usepackage{eufrak}
\usepackage{amsthm}

\def\textbf#1{\boldsymbol{#1}}

\begin{document}

\title{Resonant X-ray Scattering Study of Charge Density Wave Correlations in YBa$_2$Cu$_3$O$_{6+x}$}
\author{S.~Blanco-Canosa}
\affiliation{Max-Planck-Institut~f\"{u}r~Festk\"{o}rperforschung,
Heisenbergstr.~1, D-70569 Stuttgart, Germany}
\affiliation{Helmholtz-Zentrum Berlin f\"{u}r Materialien und Energie,
Albert-Einstein-Strasse 15, D-12489 Berlin, Germany}

\author{A.~Frano}
\affiliation{Max-Planck-Institut~f\"{u}r~Festk\"{o}rperforschung,
Heisenbergstr.~1, D-70569 Stuttgart, Germany}
\affiliation{Helmholtz-Zentrum Berlin f\"{u}r Materialien und Energie,
Albert-Einstein-Strasse 15, D-12489 Berlin, Germany}

\author{E.~Schierle}
\affiliation{Helmholtz-Zentrum Berlin f\"{u}r Materialien und Energie,
Albert-Einstein-Strasse 15, D-12489 Berlin, Germany}

\author{J.~Porras}
\affiliation{Max-Planck-Institut~f\"{u}r~Festk\"{o}rperforschung,
Heisenbergstr.~1, D-70569 Stuttgart, Germany}

\author{T.~Loew}
\affiliation{Max-Planck-Institut~f\"{u}r~Festk\"{o}rperforschung,
Heisenbergstr.~1, D-70569 Stuttgart, Germany}

\author{M.~Minola}
\affiliation{Max-Planck-Institut~f\"{u}r~Festk\"{o}rperforschung,
Heisenbergstr.~1, D-70569 Stuttgart, Germany}

\author{M.~Bluschke}
\affiliation{Max-Planck-Institut~f\"{u}r~Festk\"{o}rperforschung,
Heisenbergstr.~1, D-70569 Stuttgart, Germany}

\author{E.~Weschke}
\affiliation{Helmholtz-Zentrum Berlin f\"{u}r Materialien und Energie,
Albert-Einstein-Strasse 15, D-12489 Berlin, Germany}

\author{B.~Keimer}
\email{b.keimer@fkf.mpg.de}
\affiliation{Max-Planck-Institut~f\"{u}r~Festk\"{o}rperforschung,
Heisenbergstr.~1, D-70569 Stuttgart, Germany}

\author{M.~Le Tacon}
\email{m.letacon@fkf.mpg.de.de}
\affiliation{Max-Planck-Institut~f\"{u}r~Festk\"{o}rperforschung,
Heisenbergstr.~1, D-70569 Stuttgart, Germany}

\date{\today}

\begin{abstract}
We report the results of a comprehensive study of charge density wave (CDW) correlations in untwinned YBa$_2$Cu$_3$O$_{6+x}$ single crystals with $0.4 \leq x \leq 0.99$ using Cu-$L_3$ edge resonant x-ray scattering (RXS). Evidence of CDW formation is found for $0.45 \leq x \leq 0.93$ (hole doping levels $0.086 \lesssim p \lesssim 0.163$), but not for samples with $x \leq 0.44$ ($p \lesssim 0.084$) that exhibit incommensurate spin-density-wave order, and in slightly overdoped samples with $x = 0.99$ ($p \sim 0.19$). This suggests the presence of two proximate zero-temperature CDW critical points at $p_{c1} \sim 0.08$ and $p_{c2} \sim 0.18$. Remarkably, $p_{c2}$ is close to the doping level that is optimal for superconductivity. The CDW reflections are observed at incommensurate in-plane wave vectors ($\delta_a$, 0) and (0, $\delta_b$) with $\delta_a \lesssim \delta_b$. Both $\delta_a$ and $\delta_b$ decrease linearly with increasing doping, in agreement with recent reports on Bi-based high-$T_c$ superconductors, but in sharp contrast to the behavior of the La$_{2-x}$(Ba,Sr)$_x$CuO$_4$ family.  The CDW intensity and correlation length  exhibit maxima at $p \sim 0.12$, coincident with a plateau in the superconducting transition temperature $T_c$. The onset temperature of the CDW reflections depends non-monotonically on $p$, with a maximum of $\sim 160$ K for $p \sim 0.12$. The RXS reflections exhibit a uniaxial intensity anisotropy. Whereas in strongly underdoped samples the reflections at ($\delta_a$, 0) are much weaker than those at (0, $\delta_b$), the anisotropy is minimal for $p \sim 0.12$, and reversed close to optimal doping. We further observe a depression of CDW correlations upon cooling below $T_c$, and (for samples with $p \geq 0.09$) an enhancement of the signal when an external magnetic field up to 6 T is applied in the superconducting state. For samples with $p \sim 0.08$, where prior work has revealed a field-enhancement of incommensurate magnetic order, the RXS signal is field-independent. This supports a previously suggested scenario in which incommensurate charge and spin orders compete against each other, in addition to individually competing against superconductivity (Blanco-Canosa {\it et al.}, Phys. Rev. Lett. {\bf 110}, 187001 (2013)). We discuss the relationship of these results to prior observations of ``stripe'' order in La$_{2-x}$(Ba,Sr)$_x$CuO$_4$, the ``pseudogap'' phenomenon, superconducting fluctuations, and quantum oscillations, as well as their implications for the mechanism of high-temperature superconductivity.
\end{abstract}

\pacs{74.20.Rp, 74.25.Gz, 74.25.Kc, 74.72.Bk}


\maketitle

\section{Introduction}

The interplay between spin and charge degrees of freedom in materials with strongly correlated electrons generates complex phase diagram in which the balance between various competing phases can be tuned through parameters such as temperature, doping, pressure, and magnetic field. In the layered cuprates, removing electrons from the undoped, Mott-insulating CuO$_2$ planes suppresses long-range antiferromagnetic order and gives rise to high temperature superconductivity~\cite{Lee_RMP06}. For large concentrations of holes, $p$, per Cu ion, superconductivity disappears, and the emerging metallic state is amenable to a description by the Fermi-liquid theory. In the underdoped regime bridging the Mott-insulating and the fully developed superconducting states, however, the physical properties of the cuprates indicate the breakdown of conventional Fermi-liquid models.~\cite{Timusk_RPP1999}

Research on the origin of the ``non-Fermi liquid'' behavior of the underdoped cuprates has uncovered evidence of charge-ordering phenomena in the CuO$_2$ planes \cite{Tranquada_Nature1995,Fink_PRB2009,Fink_PRB2011,Hucker_PRB2011,Huecker_PRB2012,Kiveslon_RMP2003,Vershinin_Science04,Abbamonte_NAtPhys2005,Koshaka_Science2007,Muschler_EPJ2010,Wilkins_PRB2011,Hoffman_Science02_1,Hanaguri_Nature04,Wise_NaturePhysics2008, Ma_PRL2008,Parker_Nature2010,Lawler_Nature2010,Wu_Natcom2013,Wu_Nature2011,Ghiringhelli_Science2012,Achkar_PRL2012,Chang_NaturePhysics2012,LeBoeuf_NaturePhysics2013,Blackburn_PRL2013,Blanco_PRL2013,
Thampy_PRB2013,Blackburn_PRB2013,Bakr_PRB2013,letacon_NaturePhysics2014,Comin_Science2014,daSilvaNeto_Science2014, Hashimoto_PRB2014, Doiron_Nature2007, Sebastian_RPP2012, Doiron-Leyraud_PRX2013}.
The initial experimental evidence for charge order was obtained in the ``214'' family [La$_{2-x}$Ba$_x$CuO$_4$ and La$_{1.8-x}$(Nd,Eu)$_{0.2}$Sr$_x$CuO$_4$] where it was found to be intimately linked to doping-induced incommensurate spin correlations~\cite{Tranquada_Nature1995, Fink_PRB2009,Fink_PRB2011,Hucker_PRB2011,Huecker_PRB2012,Wilkins_PRB2011}. For $p \sim 1/8$, uniaxial ``stripe'' domains with approximately commensurate periodicity ($\sim 4$ lattice constants $a$ for charge, and $\sim 8 a$ for spin degrees of freedom) and correlation lengths up to several tens of unit cells suppress the development of superconductivity. The influence of static or fluctuating stripe domains on the fermiology of underdoped cuprates has been extensively discussed~\cite{Zaanen_PRB1989,Kato_JPSJ1990,Votja_AdvPhys2009,Kiveslon_RMP2003}.  However, disorder introduced by the randomly placed Sr/Ba donor sites remains a significant impediment to a full understanding of this key issue, since it has for instance precluded the observation of coherent quantum transport phenomena in the 214 system.

In YBa$_2$Cu$_3$O$_{6+x}$ (hereafter YBCO$_{6+x}$) and related ``123'' compounds, doping-induced disorder is significantly reduced\cite{Bobroff_PRL02} because the oxygen dopant atoms are arranged in CuO chains stacked between the CuO$_2$ layers. Depending on the oxygen content, $x$, the chains form different ordering patterns ranging from the ``ortho-II'' structure for $x \sim 0.5$, where full and empty chains alternate, to the ``ortho-I'' structure for $x \sim 1$, where all oxygen positions in the CuO chain layer are occupied.~~\cite{Zimmermann_PRB2003} In ortho-II ordered YBCO$_{6+x}$ crystals with $x \sim 0.5$ ($p \sim 0.1$), quantum oscillations have been observed in both transport and thermodynamic experiments in magnetic fields sufficient to obliterate superconducting long-range order~\cite{Doiron_Nature2007,Sebastian_RPP2012,Sebastian_PRL2012,Laliberte_NatCom2011,Bangura_PRL2008}. Nuclear magnetic resonance (NMR) experiments motivated in part by these results revealed a magnetic-field-induced modulation of the charge density in underdoped YBCO$_{6+x}$, without any signature of static magnetism~\cite{Wu_Natcom2013,Wu_Nature2011}.

Subsequent resonant ~\cite{Ghiringhelli_Science2012,Achkar_PRL2012,Blanco_PRL2013,Thampy_PRB2013} and non-resonant ~\cite{Chang_NaturePhysics2012,Blackburn_PRL2013,Blackburn_PRB2013} x-ray scattering experiments demonstrated static \cite{letacon_NaturePhysics2014} CDW order with domain sizes up to $\sim 20$ unit cells even in the absence of magnetic fields. The temperature and magnetic field dependence of the x-ray intensity implies a competition between CDW formation and superconductivity in YBCO$_{6+x}$. The x-ray studies determined the periodicity of the charge-ordered state, which turned out to be incommensurate with the underlying lattice. The CDW wavevector is along the Cu-O bond directions in the CuO$_2$ planes, with $q_{CDW} =(\delta_{a},0,0.5)$, $(0,\delta_{b},0.5)$ and $\delta_{a} \lesssim \delta_{b} \sim 0.3$. [We quote the wavevector coordinates $q = (h,k,l)$ in reciprocal lattice units (r.l.u.) based on an orthorhombic unit cell where the $c$-axis is perpendicular to the CuO$_2$ planes, and the $b$-axis is parallel to the CuO chains.], and is consistent with model calculations that attribute the small Fermi surface pockets seen in the quantum oscillation experiments to a Fermi-surface reconstruction triggered by bi-axial CDW order.~\cite{Harrison_PRL2011}

Resonant x-ray scattering (RXS) experiments have also revealed evidence of CDW order in  Bi$_2$Sr$_2$CuO$_{6+\delta}$ and Bi$_2$Sr$_2$CaCu$_2$O$_{8+\delta}$, ~\cite{Comin_Science2014,daSilvaNeto_Science2014, Hashimoto_PRB2014} in good agreement with prior results of surface-sensitive scanning tunneling spectroscopy measurements.~\cite{Hanaguri_Nature04,Ma_PRL2008,Wise_NaturePhysics2008,Parker_Nature2010,Hoffman_Science02_1,Lawler_Nature2010} Together with recent data on HgBa$_2$CuO$_{4+\delta}$ that also indicate CDW correlations, \cite{Tabis} these observations demonstrate that the CDW is a generic feature of the underdoped cuprates. An important open question is the relationship between the ``pseudogap'', another phenomenon that is ubiquitous in underdoped cuprates, and the gap associated with CDW formation. A comparative RXS and angle-resolved photoemission spectroscopy (ARPES) study of Bi$_2$Sr$_2$CuO$_{6+\delta}$ has begun to address this question by showing that the CDW wavevector matches the distance between the tips of the ungapped segments (``Fermi arcs'') of the quasi-two-dimensional Fermi surface~\cite{Comin_Science2014}.

Most of these studies (except those on the 214 system) have been carried out over a limited range of doping levels. In YBCO$_{6+x}$, CDW correlations have been observed between $p \sim 0.1$~\cite{Ghiringhelli_Science2012, Blackburn_PRL2013,Blanco_PRL2013,Thampy_PRB2013} and $p \sim 0.13$~\cite{Achkar_PRL2012}, where static magnetism is absent and the magnetic response is fully gapped~\cite{Dai_Science1999,Fong_PRB2000}. In this range of $p$, the incommensurability $\delta$ decreases with increasing $p$, as expected based on models that link the CDW to the Fermi surface, but in contrast to the 214 materials which exhibit the opposite trend. For ortho-II ordered YBCO$_{6+x}$ with $p \sim 0.1$, the charge modulation is weaker than the one observed at higher doping levels, and the CDW amplitude along the $h$-direction of reciprocal space is considerably lower than along $k$ ~\cite{Blackburn_PRL2013,Blanco_PRL2013}. For lower $p$, static incommensurate magnetic short-range order with propagation vector along $h$ has been observed~\cite{Fujita_JPSJ2012,Haug_PRL2009,Haug_NJP2010}, but evidence for a modulation of the charge density has not been reported~\cite{Ghiringhelli_Science2012}.

In order to provide further insight into the relationship between CDW correlations, quantum oscillations, the pseudogap, and superconductivity, we have undertaken a comprehensive RXS study of the doping, temperature, and magnetic field dependence of the CDW in YBCO$_{6+x}$ covering doping levels ranging from $p \sim 0.07$ ($x = 0.4$) to $p \sim 0.19$ ($x = 0.99$). The results complement and extend prior RXS work in a more limited range of $p$. Important new results include the linear doping dependence of $q_{CDW}$ over the entire range of $p$ where CDW are observable by RXS, the systematic evolution of the in-plane anisotropy of the CDW, and the discovery of CDW correlations up to (but not beyond) optimal doping ($p \sim 0.16$). The CDW quantum critical point near optimal doping indicated by these results may have important implications for the mechanism of high-temperature superconductivity.

\section{Experimental details}

\subsection{Single crystals}

\begin{figure}
\includegraphics[width=1.0\linewidth]{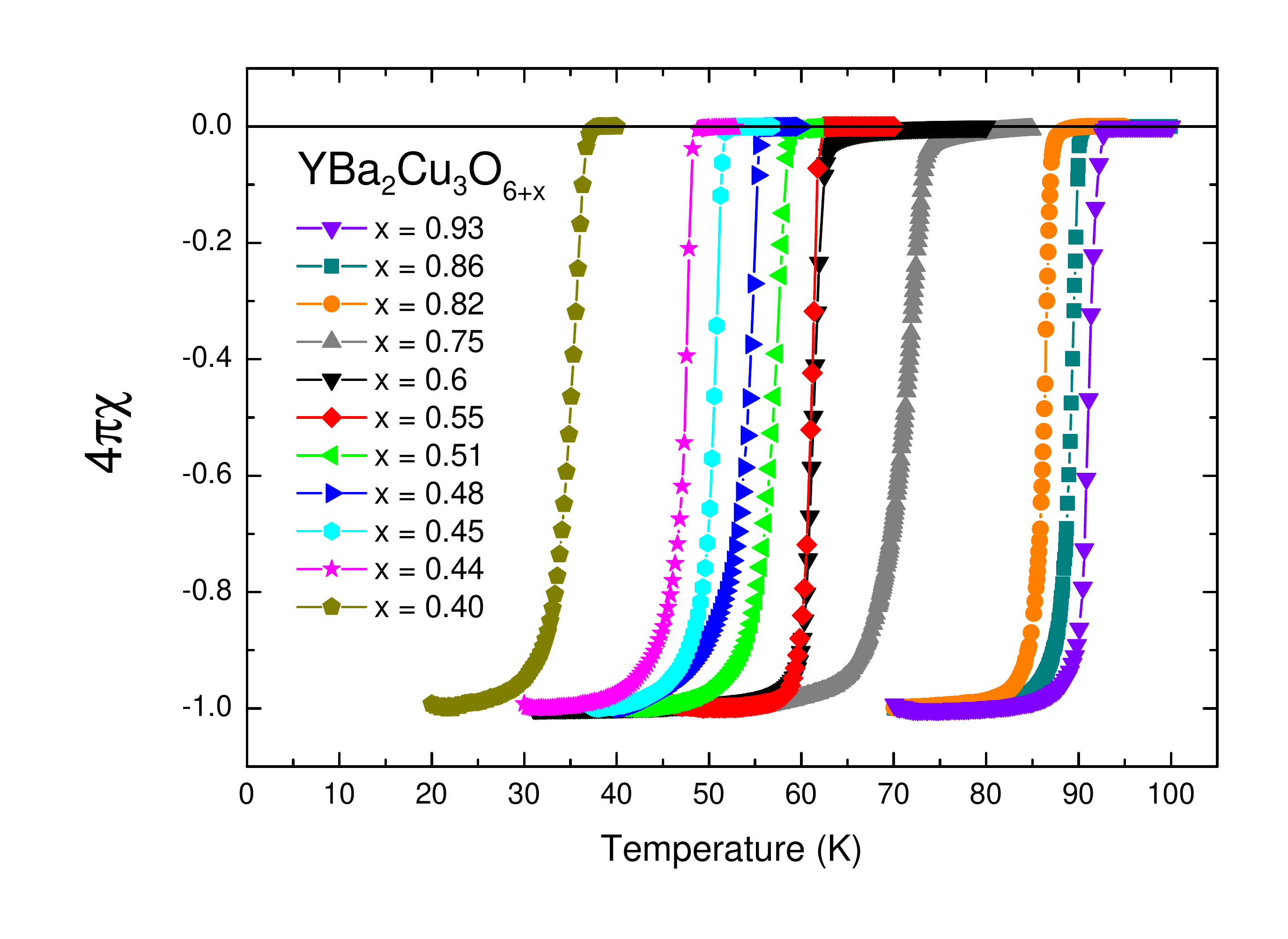}
\caption{(Color online) Magnetization curves of the set of samples investigated here.
} \label{fig:Tc}
\end{figure}

\begin{table}
\begin{tabular}{|c | c | c | c | c | c| c | c | c |}
 \hline
Sample & Structure & $T_c$ (K)& $c$-axis (\AA) & $p$  \\
\hline \hline
YBCO$_{6.40}$ & O-II & 35 & 11.771 & 0.072\\ \hline
YBCO$_{6.44}$ & O-II & 47.4 & 11.760 & 0.084\\ \hline
YBCO$_{6.45}$ & O-II & 50.5 & 11.758 & 0.086\\ \hline
YBCO$_{6.48}$ & O-II & 54.2 & 11.752 & 0.092\\ \hline
YBCO$_{6.51}$ & O-II & 57 & 11.745 & 0.099\\ \hline
YBCO$_{6.55}$ & O-II & 61 & 11.731 & 0.114\\ \hline
YBCO$_{6.6}$ & O-VIII & 61 & 11.72 & 0.12\\ \hline
YBCO$_{6.75}$ & O-III & 71 & 11.7156 & 0.134\\ \hline
YBCO$_{6.82}$ & O-III & 86.5 & 11.706 & 0.148\\ \hline
YBCO$_{6.86}$ & O-I & 89 & 11.703 & 0.152\\ \hline
YBCO$_{6.93}$ & O-I & 91 & 11.6969 & 0.163\\ \hline
YBCO$_{6.99}$ & O-I & 90 & 11.6835 & 0.189\\ \hline
\hline
\end{tabular}
\caption{List of the  YBa$_2$Cu$_3$O$_{6+x}$ crystals investigated by RXS. The structural arrangement of oxygen donor atoms is labeled O-II for ortho-II, etc. ~\cite{Zimmermann_PRB2003,Strempfer_PRL2004} The superconducting transition temperature $T_c$ was determined by magnetometry. The room temperature out-of-plane lattice
parameter, $c$, was determined by hard x-ray diffraction. From this value, the hole doping
level $p$ per planar Cu ion was extracted following Ref.~\onlinecite{Liang_PRB2006}.
}
\end{table}

YBCO$_{6+x}$ single crystals were synthesized using a flux method as described in previous reports \cite{Hinkov_Nature}. In addition to the crystals with oxygen contents $x = 0.55$ and 0.6 previously studied by RXS \cite{Blanco_PRL2013, Ghiringhelli_Science2012, Thampy_PRB2013}, we present results obtained on single crystals with both lower $x$ ranging from 0.40 ($p \sim 0.07$) to 0.51 ($p \sim 0.1$), and higher $x$ between 0.75 ($p \sim 0.14$) and 0.99 ($p \sim 0.19$). The oxygen content was controlled by annealing under well-defined oxygen partial pressure
All samples were mechanically detwinned by heating under uniaxial stress.
Subsequently, the ortho-II and ortho-III phase crystals have been annealed below their corresponding superstructure ordering temperatures following refs.~\cite{Zimmermann_PRB2003,Liang_PRB2006} for an improved CuO chain ordering. The $c$-axis lattice parameters of the samples, determined from hard x-ray diffraction measurements, were used to obtain their hole-doping levels~\cite{Lindemer_JACS1989, Liang_PRB2006}, and the superconducting $T_c$ of the crystals was determined from the midpoint of the transition measured in a VSM SQUID magnetometer (Fig.~\ref{fig:Tc}).

Our hard x-ray diffraction measurements of the ortho-II type of oxygen superstructure, comprising alternating full and empty CuO chains, indicate correlation lengths $\xi_a$ between $\sim$ 10 \AA~and $\sim$ 100 \AA ~\cite{Blanco_PRL2013} for YBCO$_{6+x}$ with 0.4 $\leq x \leq$ 0.55.
Due to an unfavorable scattering geometry, we cannot determine this correlation length directly from soft x-ray scattering measurements. A qualitative picture can, however, be drawn from the doping dependence of the width of the superstructure peak measured at the L$_3$ resonance of chain Cu at $q = (0.5, 0, l)$~\cite{Blanco_PRL2013,Achkar_PRL2012}; note that in our rocking scans that cover the range 0.45-0.55 for $h$, $l$ varies between $\sim 1$ and 0.5 (see Section~\ref{sec:resonant}). 
The YBCO$_{6.6}$ crystal showed ortho-VIII correlations, and ortho-III correlations were observed in the crystals with higher oxygen contents, again in agreement with prior work~\cite{Zimmermann_PRB2003,Strempfer_PRL2004}. Crystals with $x \geq 0.79$ only showed ortho-I correlations.

\begin{figure}
\includegraphics[width=1.0\linewidth]{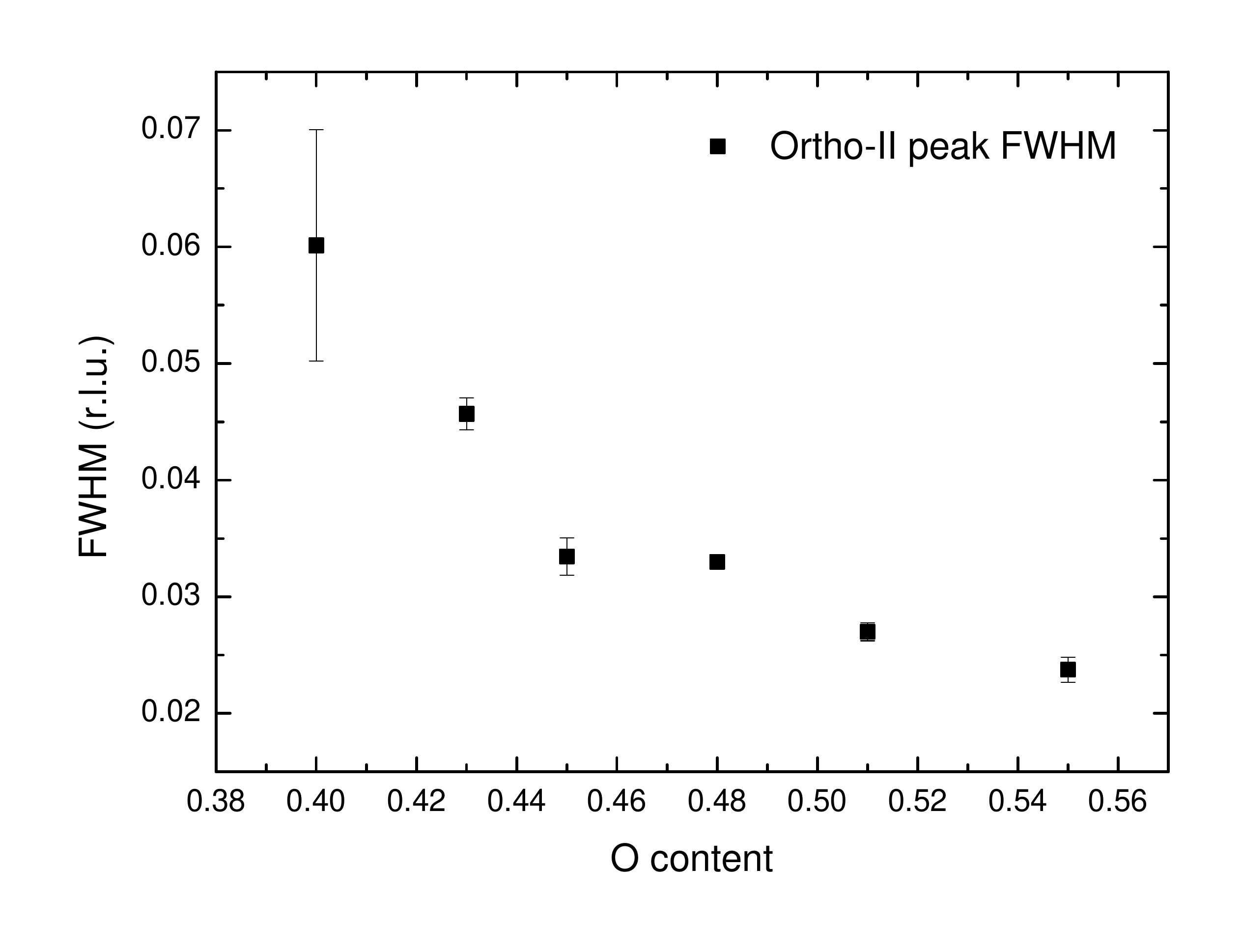}
\caption{(Color online) Full-width-at-half-maximum of the ortho-II superstructure peak measured along the at $q=(0.5, 0, l)$ with photon energy tuned to the $L_3$ absorption edge of the chain Cu atoms as a function of the oxygen content $x$ in YBCO$_{6+x}$.
}
\label{fig:orthoII}
\end{figure}

\subsection{Resonant x-ray scattering}
\label{sec:resonant}

In this section, we recall some basis features of resonant soft x-ray scattering (RXS). For more details, see the recent review in Ref.~\onlinecite{Fink_RPP2013}. The resonant scattering intensity $I$($\omega$) depends on the incoming photon energy, $\hbar \omega$, and can be described as

\begin{equation}
 I(\omega) = \bigg | \sum_{\substack{
  n
  }}
 \mathrm{e}^{\imath q \cdot R_n}(\textbf{\epsilon}'^\ast \cdot F_n(\omega) \cdot \textbf{\epsilon})\bigg |^2
\label{eq:ResonantIntensity}
\end{equation}

where $q$ denotes the scattering vector, $R_n$ is the position of the $n^{th}$ atom, $\epsilon$ ($\epsilon'$) is the incoming (outgoing) polarization, and $F_n(\omega)$ is the energy dependent scattering tensor (also known as form factor) for each atomic species. $F_n$ reflects both charge and magnetic degrees of freedom.




By tuning the incident photon energy to a specific x-ray absorption edge, the atomic structure factor $F_n$ is strongly enhanced. RXS experiments are therefore directly sensitive to the valence electron states, as compared to other techniques such as nuclear neutron or hard x-ray scattering which probe the lattice displacements induced by the modulation of the valence electron density.

The YBCO structure contains three Cu ions per unit cell, two in the CuO$_2$ planes, with a 3d$^{9}$ oxidation state with an electron in the 3d$_{x^{2}-y^{2}}$ orbital,
and one in the CuO chains, with valence states changing from 3d$^{10}$ to 3d$^{9}$ as the excess of oxygen $x$ varies from 0 to 1. These yield several final states for the XAS process, that has been studied in details in the literature~\cite{Nuecker_PRB1995,Hawthorn_PRB2011}.
In this work, except for the characterization of the chain order mentioned in the previous section, all the measurements have been performed at the Cu-L$_3$ edge of planar Cu ($\sim$ 931.5 eV).

Zero-field RXS measurements were performed in the UHV diffractometer at the UE46-PGM1 beamline of the Helmholtz-Zentrum Berlin at BESSY-II, with incoming light polarization perpendicular to the scattering plane. Magnetic field dependent measurements (up to $H = 6$ T) were performed at the same beamline. The field was applied at an angle of 11.5$^{\circ}$ to the $c$-axis, nearly perpendicular to the CuO$_2$ planes. The data were not corrected for self-absorption.
The background measured in the magnet chamber was found to be independent of the applied magnetic field.

The crystals were aligned with the CuO$_2$ planes perpendicular to the scattering plane. The data presented hereafter consist of rocking scans taken around the CDW peak positions in the $(h,0,l)$ and $(0,k,l)$ planes of the reciprocal space. Across the region of interest (0.25 $\lesssim h,k\lesssim$ 0.35), $l$ between 1.40 and 1.55,  which is close to the half-integer $l$-value that maximizes the scattering intensity of the CDW peak~\cite{Chang_NaturePhysics2012, letacon_NaturePhysics2014}.

\section{Results}

\subsection{Doping dependence}

\begin{figure}
\includegraphics[width=1.0\linewidth]{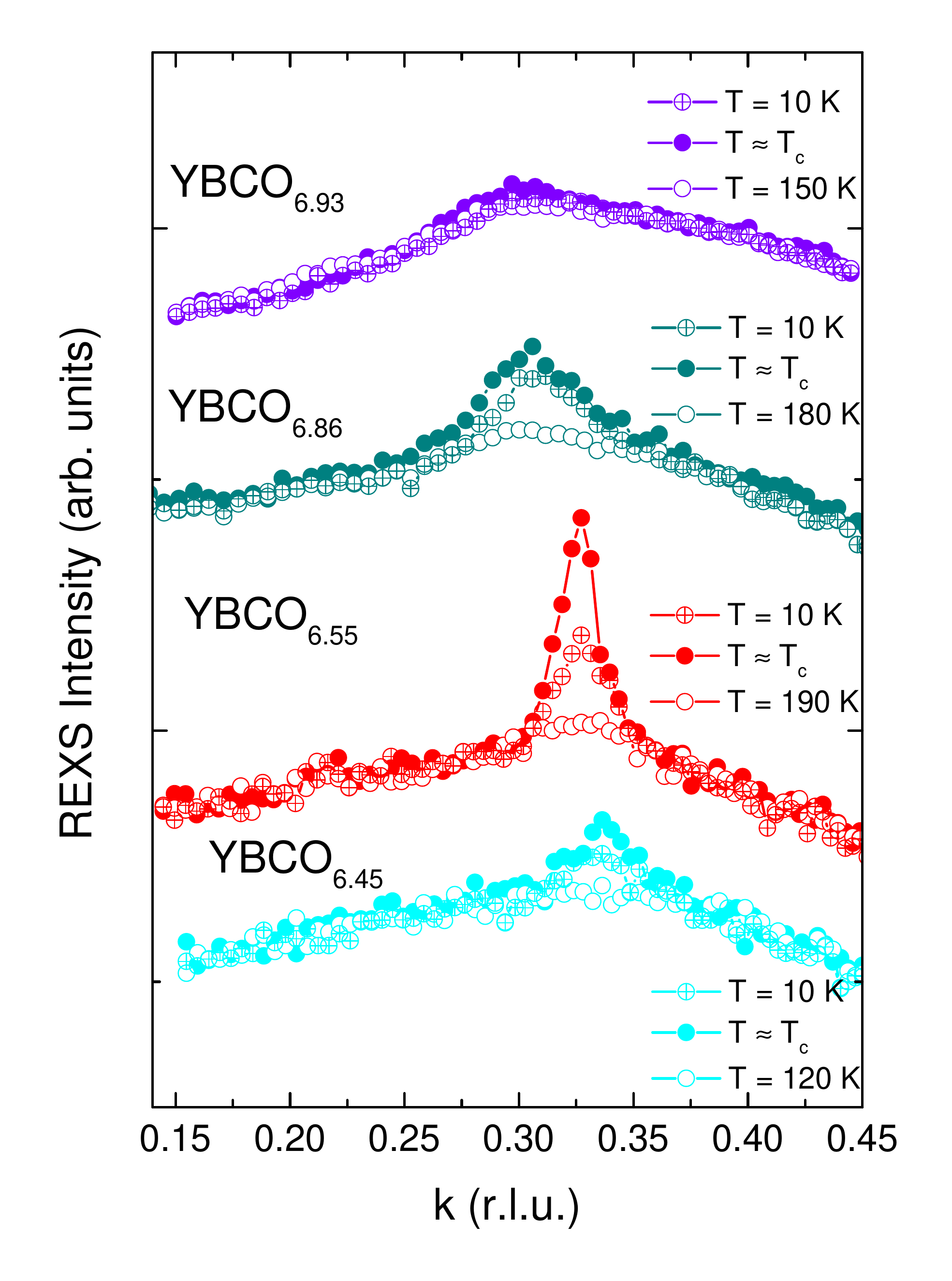}
\caption{(Color online) Raw data measured along the (0,1,0) direction for YBa$_2$Cu$_3$O$_{6+x}$ samples with $x = 0.45$, 0.55, 0.86, and 0.93 close to their respective $T_c$ (full symbols) and above the onset of the CDW signal (empty symbols).}
\label{fig:Raw_with_backgnd}
\end{figure}

\begin{figure}
\includegraphics[width=1.0\linewidth]{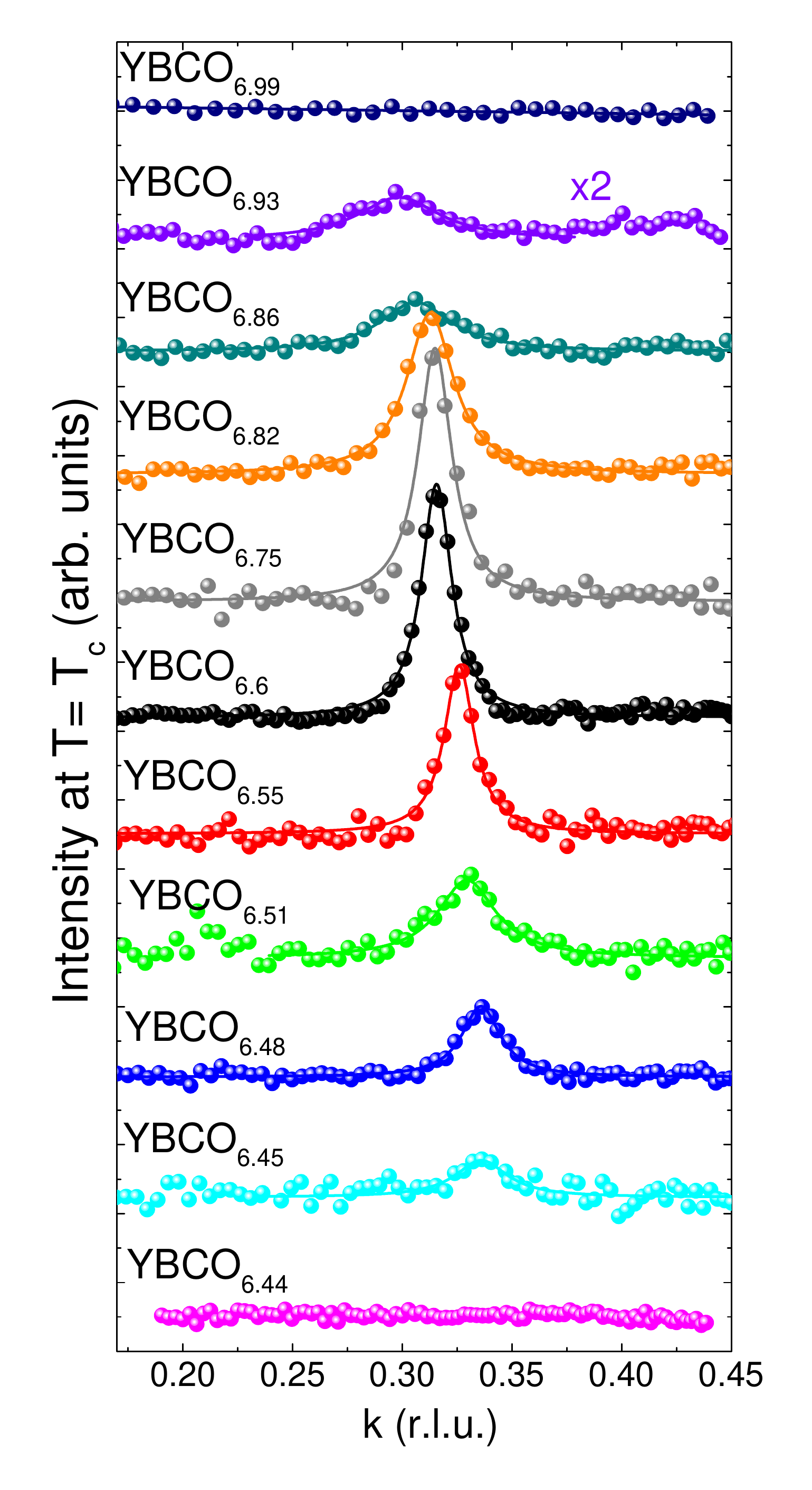}
\caption{(Color online) Background-subtracted RXS intensity measured along the (0, 1, 0) direction for a set of YBa$_2$Cu$_3$O$_{6+x}$ samples with $0.44 < x < 0.99$. Solid lines are the results of fits to Lorentzian profiles.
}
\label{fig:REXS_DopingDep}
\end{figure}

Figure~\ref{fig:Raw_with_backgnd} presents raw data taken on different YBCO$_{6+x}$ crystals at their respective $T_c$, together with background scans at higher temperatures. Many of our key results are already apparent in the unprocessed scans. As reported in Ref.~\onlinecite{Blanco_PRL2013}, a superstructure peak associated with CDW correlations is observed in the RXS spectra of YBCO$_{6.55}$ at $q_{CDW} =(0,0.326, l)$. With both increasing and decreasing oxygen content, CDW peaks remain clearly visible on top a temperature independent background, which was determined by measuring the RXS signal at higher temperatures above which it remains temperature independent. This onset temperature ranges from $\sim  110$ to 160 K depending on the doping level (see below). Note that an extremely broad, temperature independent peak centered at approximately the same position remains visible even above the onset temperature determined in this way.

Special care was taken to perform all the measurements presented in this paper under comparable experimental conditions. Although quantitative comparisons of absolute RXS intensities from sample to sample remain difficult, especially since details of the oxygen order can significantly affect the intensity of the CDW peak~\cite{Hawthorn_disorder}, a pronounced intensity maximum for $p \sim 0.12$ can be clearly identified on a qualitative level. The temperature, magnetic field, and wave vector dependence of the RXS intensity for individual samples as well as the $p$-dependence of the peak position and width are unaffected by sample-to-sample intensity variations and can be accurately extracted from the data in Fig. \ref{fig:Raw_with_backgnd}.

Figure~\ref{fig:REXS_DopingDep} displays background-subtracted data obtained for representative samples with oxygen concentrations ranging from $x = 0.44$ to 0.99 at their respective $T_c$. Signatures of CDW formation are not found for oxygen contents lower than $x = 0.45$ ($p = 0.086$) and in fully oxygenated YBCO$_{6.99}$ ($p = 0.189$). At all other doping levels, temperature dependent CDW correlations can be identified, with a pronounced intensity maximum for $p \sim 0.12$. We will henceforth refer to the doping range $0.08 \lesssim p \lesssim 0.18$ where CDW correlations are observable by RXS as the ``CDW stability range''. Since the CDW correlation length always remains finite, however, these data do not imply thermodynamic stability of the CDW.

\subsection{Anisotropy}

\begin{figure}
\includegraphics[width=1.0\linewidth]{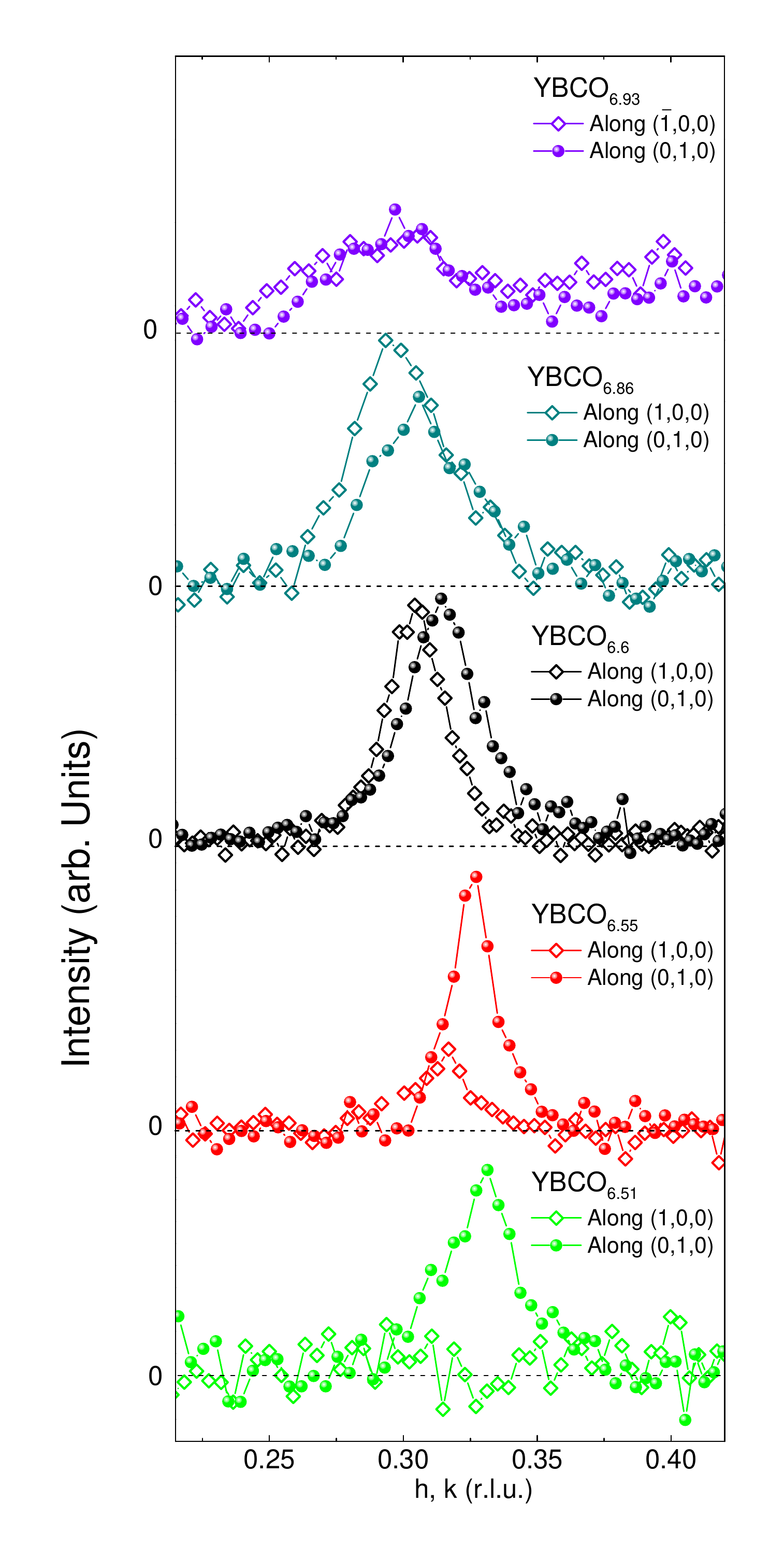}
\caption{(Color online) In-plane anisotropy of the background-subtracted RXS signal in YBa$_2$Cu$_3$O$_{6.86}$, YBa$_2$Cu$_3$O$_{6.6}$, YBa$_2$Cu$_3$O$_{6.55}$, and YBa$_2$Cu$_3$O$_{6.51}$. Full and empty symbols stand for data taken along the (0,1,0) and (1,0,0) directions, respectively.}
\label{fig:REXS_a_vs_b}
\end{figure}

\begin{figure}
\includegraphics[width=1.0\linewidth]{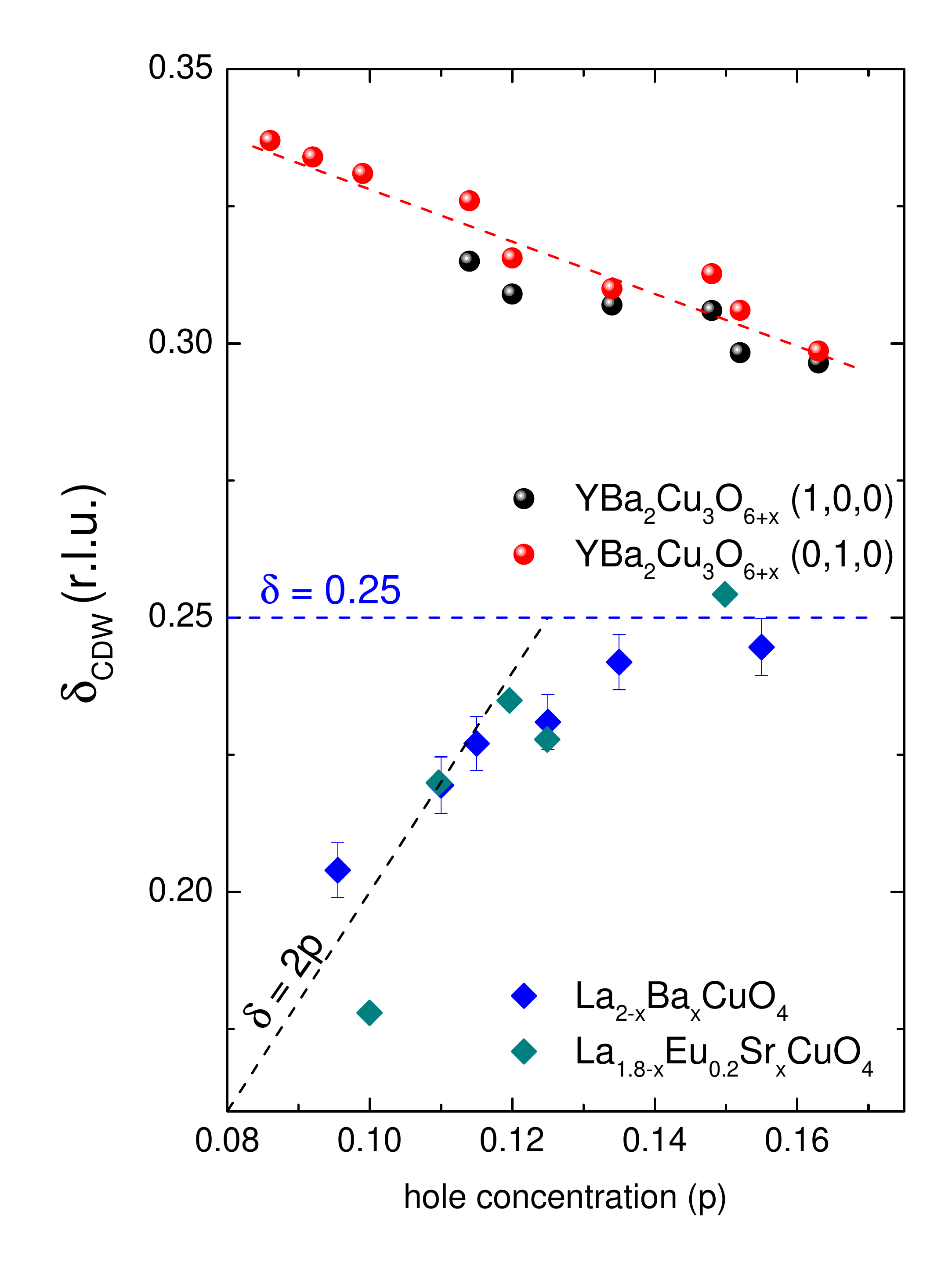}
\caption{(Color online) Doping dependence of the CDW wavevector in YBa$_2$Cu$_3$O$_{6+x}$ compared to the wave vector characterizing charge order in the ``striped'' state of La$_{2-x}$Ba$_x$CuO$_4$ (Ref.~\onlinecite{Hucker_PRB2011}) and La$_{1.8-x}$Eu$_{0.2}$Ba$_x$CuO$_4$ (Ref.~\onlinecite{Fink_PRB2011}).}
\label{fig:QCDW}
\end{figure}

So far, we have focused on the CDW peak along the (0,1,0) ($k$)-direction. As already noted in Refs.~\onlinecite{Blackburn_PRL2013,Blanco_PRL2013}, the CDW is highly anisotropic in the ortho-II ordered sample YBCO$_{6.55}$, where the intensity is strongly reduced along the (1,0,0) ($h$)-direction. In the samples with $x < 0.55$ investigated here, the CDW peak is only observed along $k$. This is illustrated in Fig.~\ref{fig:REXS_a_vs_b} where background-subtracted data at $T = T_c$ are shown along both $h$ and $k$. At higher doping levels, the CDW is much more isotropic, as previously reported in YBCO$_{6.6}$~\cite{Ghiringhelli_Science2012} and YBCO$_{6.75}$~\cite{Achkar_PRL2012}. For $x = 0.86$, we now find that the peak along $h$ is slightly more intense than the one along $k$, so that the anisotropy is reversed compared to the samples with $x < 0.6$.

Figure~\ref{fig:REXS_a_vs_b} also shows that at each doping level where the signatures of charge modulations are seen in both directions, the incommensurability along $h$ is always slightly smaller than along $k$, as pointed out in Refs.~\onlinecite{Blackburn_PRL2013, Blanco_PRL2013} for a more limited set of samples. In order to extract the peak width and position, the data of Fig.~\ref{fig:REXS_DopingDep} were fitted to Lorentzian profiles. Figure~\ref{fig:QCDW}a shows a summary plot of the doping dependence of the incommensurability determined in this way over the entire CDW stability range. Both $\delta_a$ and $\delta_b$ decrease linearly with increasing doping. This behavior contrasts markedly with the one in the 214 materials shown for comparison in Fig.~\ref{fig:QCDW}b, where $\delta$ first increases with increasing $p$ and then saturates for $p \sim 1/8$.

\begin{figure}
\includegraphics[width=1.0\linewidth]{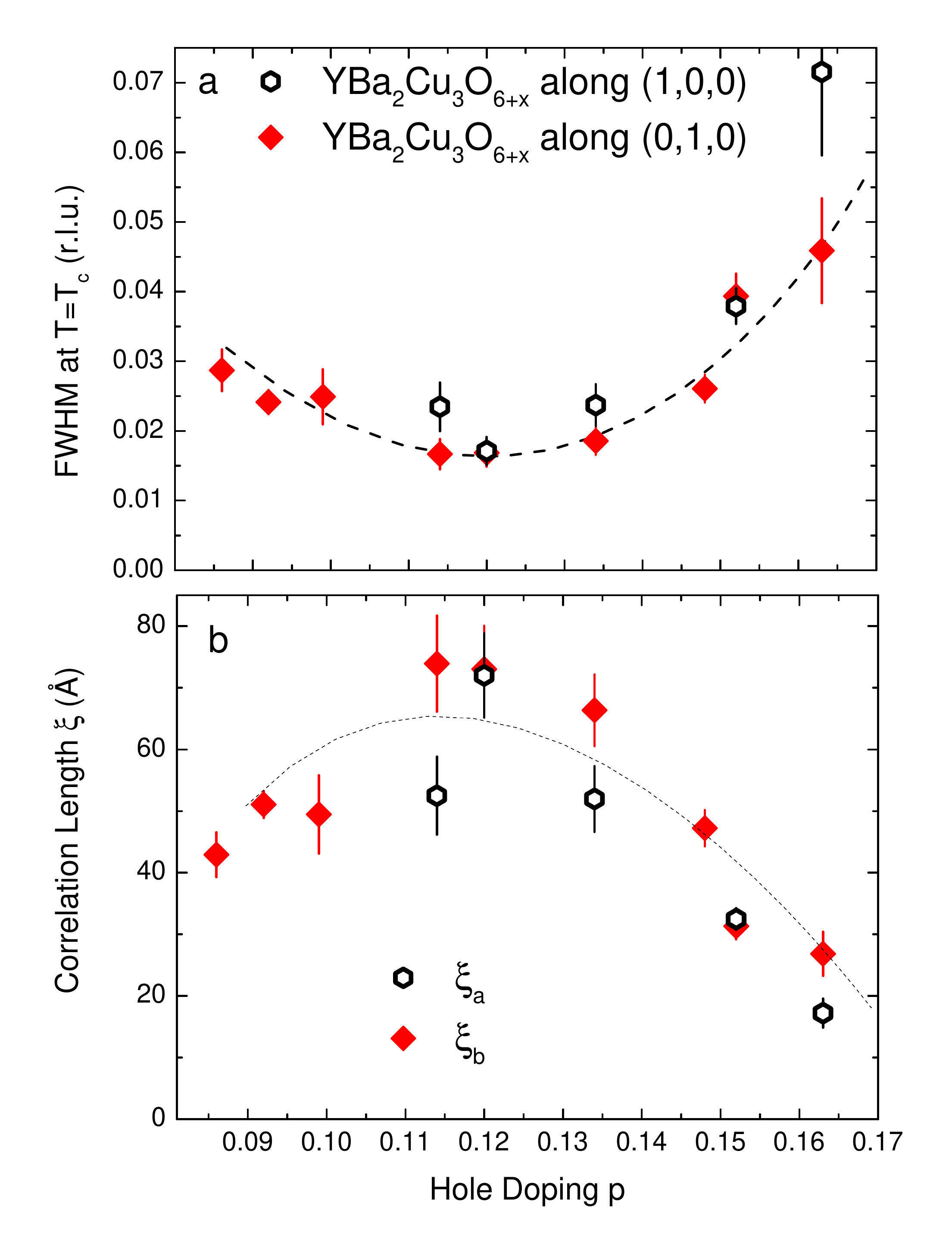}
\caption{(Color online) a) CDW peak full-width-at-half-maximum (FWHM) as function of doping, given in reciprocal lattice units (r.l.u.) and measured at $T = T_c$ in our YBa$_2$Cu$_3$O$_{6+x}$ samples. b) Corresponding correlation length, defined as $\xi_{a,b} = \frac{a,b}{\pi\times FWHM}$. Dashed lines are guides to the eyes.
}
\label{fig:FWHM_at_Tc}
\end{figure}

Figure~\ref{fig:FWHM_at_Tc}a displays the doping dependence of the Lorentzian full-width-at-half-maximum (FHWM) of the CDW peak at $T_c$ in both directions. The correlation length $\xi$ extracted from these data (Fig.~\ref{fig:FWHM_at_Tc}b) reaches a maximum of $\sim 75$ \AA (about 20 lattice spacings) for $p \sim 0.12$, mirroring the amplitude maximum inferred from the raw data in Fig.~\ref{fig:Raw_with_backgnd}. Near the end points of the CDW stability range, $\xi \sim 30$ \AA (about 8 lattice spacings), comparable to the CDW correlation lengths observed in Bi$_2$Sr$_2$CuO$_{6+\delta}$, Bi$_2$Sr$_2$CaCu$_2$O$_{8+\delta}$, and HgBa$_2$CuO$_{4+\delta}$. For the samples with the largest $\xi$, the peak widths in the $h$- and $k$-directions differ by up to $\sim 50$\%, which translates into a highly anisotropic correlation volume in the CuO$_2$ planes.

\subsection{Temperature dependence}

\begin{figure}
\includegraphics[width=1.0\linewidth]{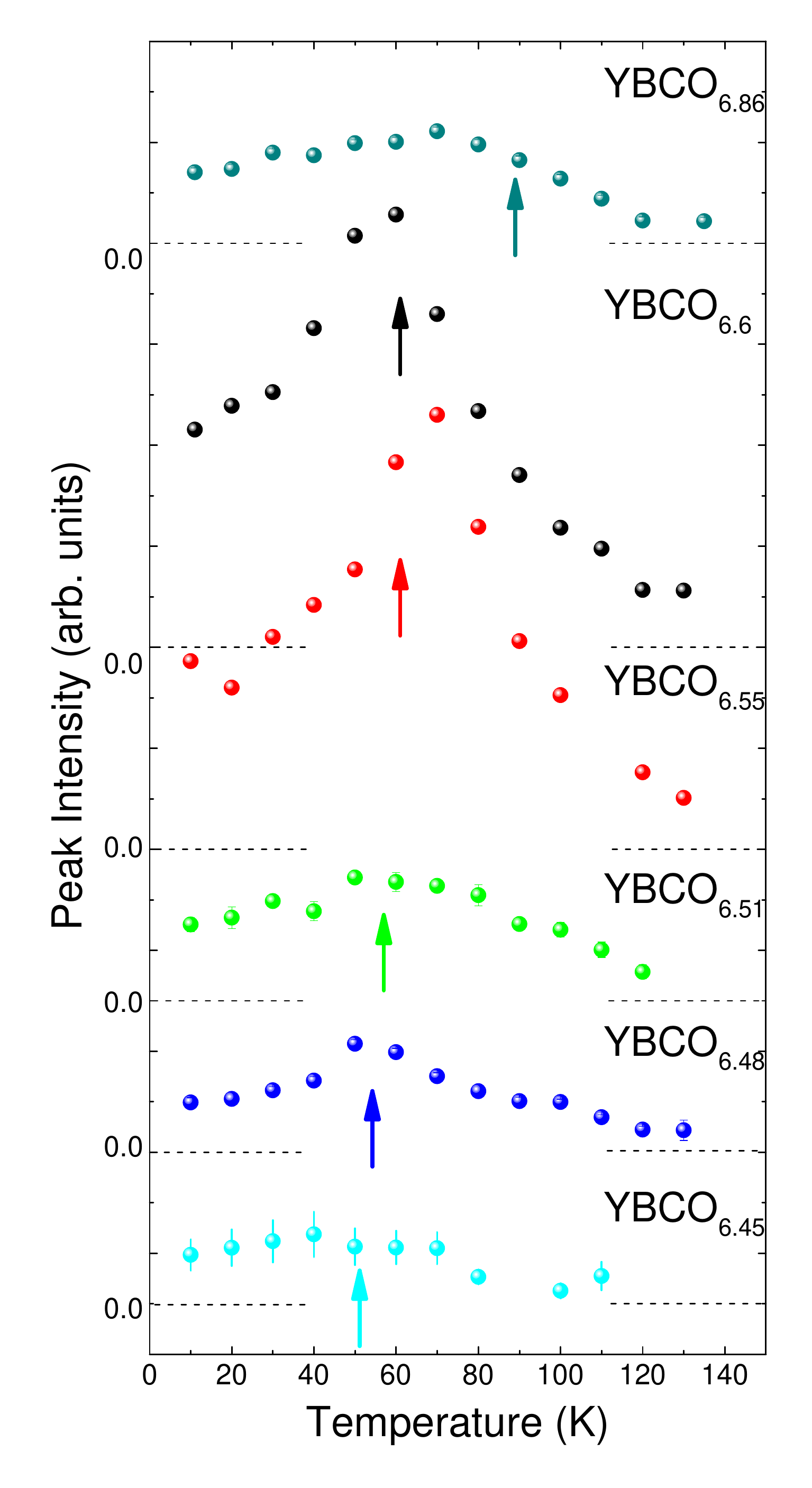}
\caption{(Color online) Temperature dependence of the CDW peak intensity for a set of YBa$_2$Cu$_3$O$_{6+x}$ samples with $0.44 < x < 0.86$ along the (0,1,0) direction. The plots have been shifted vertically for clarity, and the arrows correspond to the superconducting $T_c$.
}
\label{fig:Intensity_vs_T}
\end{figure}

\begin{figure}
\includegraphics[width=1.0\linewidth]{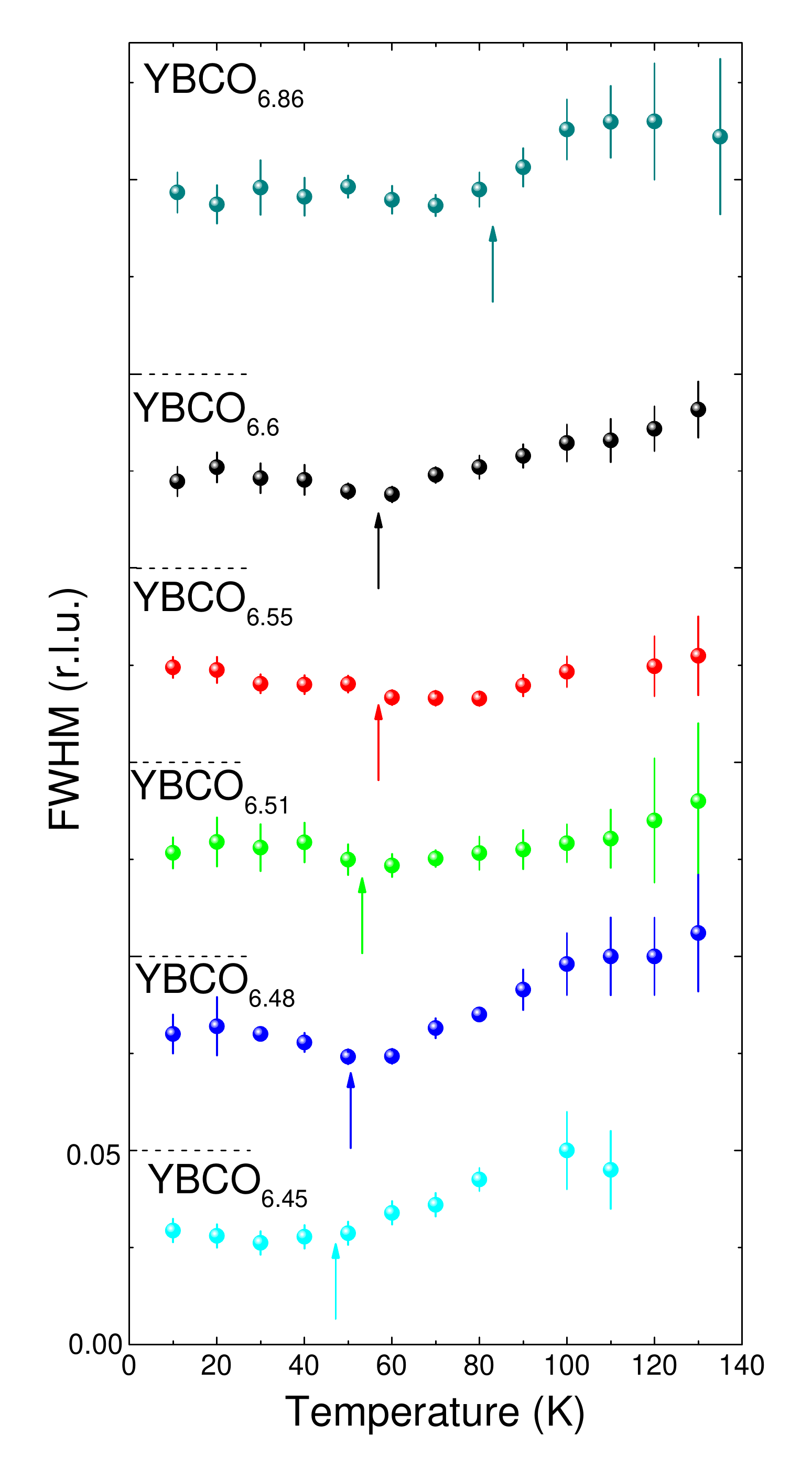}
\caption{(Color online)  Temperature dependence of the CDW peak FWHM for a set of YBa$_2$Cu$_3$O$_{6+x}$ samples with $0.44 < x < 0.86$ along the (0,1,0) direction. The plots have been shifted vertically for clarity, and the arrows correspond to the superconducting $T_c$.
}
\label{fig:FWHM_vs_T}
\end{figure}

The temperature dependence of the CDW peak intensity is plotted in Fig.~\ref{fig:Intensity_vs_T} for representative samples. In agreement with prior work, we note that the intensity is maximal around $T_c$ in all samples except the one with $x = 0.86$, where the maximum appears to be slightly below $T_c$. The superconductivity-induced intensity reduction is most pronounced for $p \sim 0.12$, and it is less marked near the end points of the CDW stability range. Fig.~\ref{fig:FWHM_vs_T} shows that the FWHM of the CDW peak follows a related trend. The peaks first become narrower upon cooling from high temperature, indicating progressive expansion of the CDW correlation volume, and then broaden again below $T_c$, reflecting the suppression of CDW order by superconductivity. Once again, this behavior is most pronounced for $p \sim 0.12$.

\begin{figure}
\includegraphics[width=1.0\linewidth]{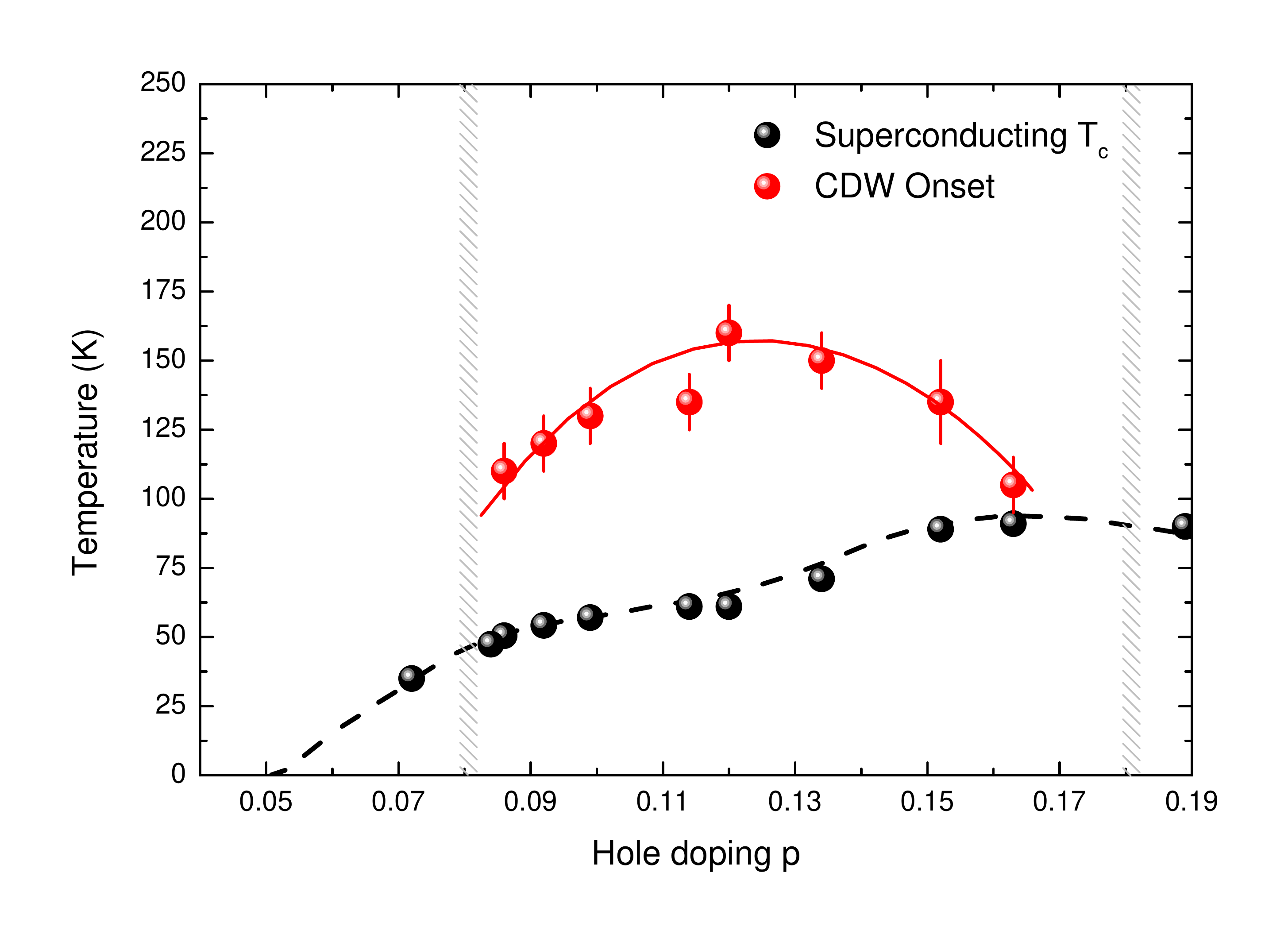}
\caption{(Color online) Doping dependence of the onset temperature of the CDW in YBa$_2$Cu$_3$O$_{6+x}$. The point at $p \sim 0.13$ has been taken from Ref.~\onlinecite{Achkar_PRL2012}.}
\label{fig:Tonset}
\end{figure}

Figure~\ref{fig:Tonset} shows the doping dependence of the onset temperature of the CDW signal, $T_{CDW}$, which depends non-monotonically on $p$. The maximum of the $T_{CDW}(p)$ ``dome'' coincides with the maxima in the CDW amplitude (Figs.~\ref{fig:Raw_with_backgnd} and ~\ref{fig:REXS_DopingDep}) and correlation length (Fig.~\ref{fig:REXS_a_vs_b}).

\subsection{Magnetic field dependence}

\begin{figure}
\includegraphics[width=1.0\linewidth]{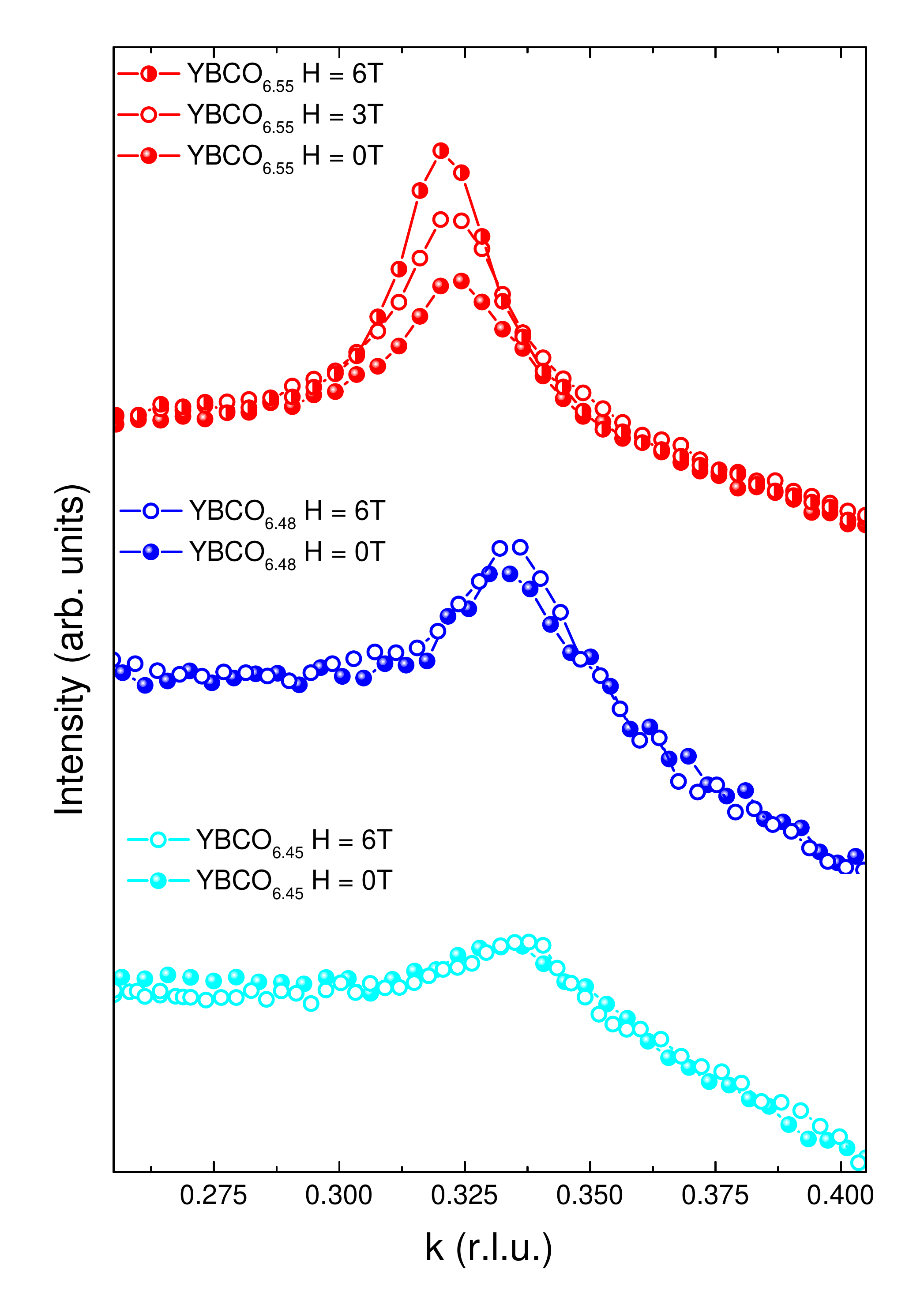}
\caption{(Color online) a) Comparison of the RXS data along the $k-$direction for YBa$_2$Cu$_3$O$_{6+x}$, with $x = 0.45$, 0.48, and 0.55 for magnetic fields $H =0$ and 6 T at $T = 4$ K. A vertical offset has been applied for clarity.
}
\label{fig:REXS_Field_data}
\end{figure}

\begin{figure}
\includegraphics[width=1.0\linewidth]{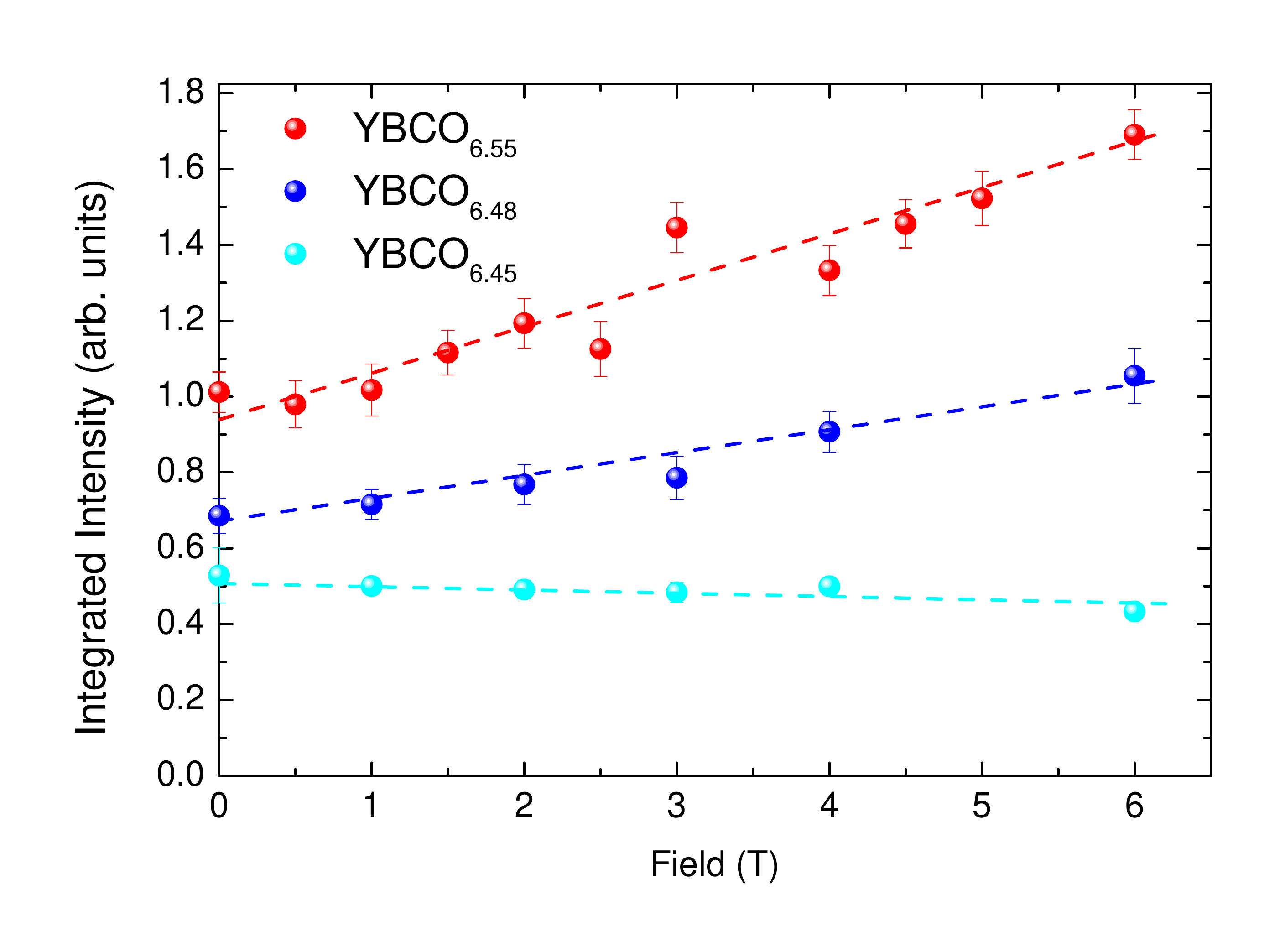}
\caption{ Magnetic field dependence of the integrated intensity of the RXS reflections (along the $k-$direction) in YBa$_2$Cu$_3$O$_{6.55}$, YBa$_2$Cu$_3$O$_{6.48}$, and YBa$_2$Cu$_3$O$_{6.55}$. Dashed lines are guides to the eye.}
\label{fig:REXS_Field_intensity}
\end{figure}

Previous hard x-ray studies reported a large enhancement of the CDW peak intensity in magnetic fields up to 17~T~\cite{Chang_NaturePhysics2012,Blackburn_PRL2013}. In the present study, the maximal magnetic field is limited to 6~T, which was already enough to observe an enhancement of the integrated intensity of the CDW peak by a factor $\sim 2$ in our YBCO$_{6.55}$ single crystal~\cite{Blanco_PRL2013}.
In Fig.~\ref{fig:REXS_Field_data}, we show analogous measurements performed on the YBCO$_{6.55}$, YBCO$_{6.48}$ and YBCO$_{6.45}$ crystals, with and without applied magnetic field at $T = 4$ K. In each case, the magnetic field was always applied at low temperature (zero field cooled procedure). In agreement with our previous study, an enhancement of the CDW peak intensity by a factor of $\sim 2$ is seen when applying a field of 6 T to the YBCO$_{6.55}$ sample. This magnetic-field induced enhancement of the CDW peak is however strongly reduced at lower $p$, and in YBCO$_{6.45}$, the effect of the field on the amplitude and width of the CDW peak is no longer discernible. The $H$-dependence of the integrated intensity in all three samples is summarized in Fig.~\ref{fig:REXS_Field_intensity}.

The magnetic field induced enhancement of the CDW peak seen in YBCO$_{6.55}$ confirms the competition between the CDW and superconductivity already apparent in Fig.~\ref{fig:Intensity_vs_T}. Due to this competition, CDW correlations are diminished in the superconducting state, but are restored when the field weakens superconductivity. The weaker field dependence observed in YBCO$_{6.48}$ and YBCO$_{6.45}$ is presumably a consequence of the competition with yet another phase, incommensurate magnetic order, which becomes the leading instability for $p < p_{c1}$~\cite{Metlitski_NJP2010}. Indeed, prior neutron scattering data have demonstrated a magnetic field induced enhancement of incommensurate spin density wave (SDW) order for a YBCO$_{6.45}$ sample, \cite{Haug_PRL2009} which mirrors the behavior of the CDW for YBCO$_{6.55}$ seen in Fig.~\ref{fig:REXS_Field_intensity}. This conclusion is supported by the fact that the linewidths of the CDW reflections in YBCO$_{6.48}$ and YBCO$_{6.45}$ remain unaffected by the magnetic field, unlike the behavior for more highly doped samples (Fig.~\ref{fig:REXS_Field_intensity}). The data presented here thus support the three-phase competition for $p \sim p_{c1}$ inferred from prior work~\cite{Blanco_PRL2013}. The effect of magnetic fields for higher $p$ has not been investigated here. Since the critical field for superconductivity is much larger around optimal doping, we do not expect any significant field effect on the CDW in the 6 T field available for this study.

\section{Discussion}
\label{sec:discussion}

We now discuss the relationship of our observations to other prominent phenomena in the underdoped cuprates. Our discussion will remain on a phenomenological level, and we will refer to the rapidly evolving theoretical literature for information about the different theoretical approaches to these issues.

\subsection{CDW quantum critical points}

The proximity of antiferromagnetic and superconducting phases in cuprates has triggered many discussions about the relevance of quantum criticality for high temperature superconductivity and associated phenomena~\cite{Sachdev_RMP2003, Metlitski_NJP2010}.
In particular, it has recently been shown that CDW states competing with superconductivity can occur next to an antiferromagnetic quantum critical point (QCP)~\cite{Efetov_NatPhys2013,Wang_arxiv,Sachdev_PRL2013,Meier_PRB2013,Meier_arxiv}.

Experimentally, based on accurate RXS data on high-quality single crystals, we have detected CDW order in YBCO$_{6+x}$ for $p_{c1} \leq p \leq p_{c2}$ with $p_{c1} \sim 0.08$ and $p_{c2} \sim 0.18$. The strong temperature dependence of the CDW peak intensity and linewidth suggest that the signal arises from fluctuations whose  divergence upon cooling is abruptly cut off below the superconducting transition temperature (Fig.~\ref{fig:Intensity_vs_T}).
Note that it has been recently argued that the shape of the temperature dependence of the peak intensity as well as the temperature range over which it exists are indicative of angular fluctuations of a multicomponent order parameter (charge order + superconductivity) rather than more conventional critical fluctuations~\cite{Hayward}.
In any event, the CDW does not exhibit genuine long-range order at any point in the phase diagram investigated in this study, and as such, the points $p_{c1}$, $T=0$ and  $p_{c2}$, $T=0$ in the two-dimensional phase diagram shown in Fig.~\ref{fig:Tonset} should not be regarded as genuine QCP. Recent transport experiments in high external magnetic fields suggest that they may instead be end points of crossover lines connected to proximate zero-temperature critical points in a three-dimensional phase diagram where the field acts as a control parameter.~\cite{Grissonnanche_NatCom2014,Ramshaw_PC}

The magnetic field weakens superconductivity and extends the divergence of the CDW correlations to lower temperatures (Fig.~\ref{fig:REXS_Field_data}). The transport  experiments indicate that this trend continues in higher fields, and that a ``naked'' CDW quantum critical point can be exposed in sufficiently large $H$. $p_{c1}$ is very close to the doping level at which quantum oscillation data have revealed an electron mass divergence pointing to a quantum critical point associated with the metal-to-insulator transition~\cite{Sebastian_PNAS2010}. Evidence has also been reported for a Lifshitz transition of the Fermi surface~\cite{LeBoeuf_PRB2011} and a maximum of the critical field for superconducting long-range order~\cite{Grissonnanche_NatCom2014} for $p \sim p_{c1}$. However, the phase behavior near $p_{c1}$ is complicated by the three-phase competition between CDW, SDW, and superconductivity, whose $p$- and $H$-evolution requires further study.

The behavior near the critical doping level $p_{c2}$ suggested by the data presented here is not affected by competition with a third phase. Recent high-field quantum oscillation data indicate an electron mass divergence for $p \rightarrow p_{c2}$ that mirrors the behavior near $p_{c1}$ and suggests that the doping-induced disappearance of the CDW is indeed associated with quantum criticality.~\cite{Ramshaw_PC} Remarkably, $p_{c2}$ is close to the doping level at which the superconducting transition temperature is maximum for $H = 0$.

\subsection{Relation to the pseudogap and superconducting fluctuations}

For the discussion of the phase diagram at nonzero $T$, we emphasize once more that the CDW onset temperature $T_{CDW}(p)$ (Fig.~\ref{fig:Tonset}) is not a thermodynamic phase boundary. Non-resonant x-ray scattering experiments with high energy resolution~\cite{letacon_NaturePhysics2014} as well as NMR experiments~\cite{wu_arxiv} rather indicate that it corresponds to the onset of static CDW short-range order nucleated by residual defects, which are present even in the highest-quality single crystals. However, the coincident maxima of the CDW onset temperature (Fig.~\ref{fig:Tonset}), amplitude (Figure~\ref{fig:REXS_DopingDep}), and correlation length (Fig.~\ref{fig:FWHM_vs_T}) at the same doping level, $p \sim 0.12$, indicate a maximum in the intrinsic strength of the CDW, independent of the nature and propensity of defects in the 123 structure. Further evidence for this line of reasoning comes from investigations of the 214 compounds, where a maximum is observed around the same doping level.\cite{Hucker_PRB2011}

Based on these considerations, we now address the relationship of the CDW and the pseudogap, another generic feature of the underdoped cuprates. Although the pseudogap onset temperature line $T^*(p)$ is still subject of debate, evidence from a variety of thermodynamic and spectroscopic probes suggest that it ends inside the superconducting dome, at a doping level that coincides with the end point of the CDW stability range, $p_{c2} \sim 0.18$, determined in the present study. \cite{Tallon_PhysicaC2001} Since the ``Fermi arc'' phenomenon is intimately tied to the pseudogap, this observation is consistent with the ``Fermi arc nesting'' scenario for CDW formation proposed by Comin \textit{et al.}~\cite{Comin_Science2014}. For $p < p_{c2}$, both $T_{CDW}(p)$ (Fig.~\ref{fig:Tonset}) and $T^*(p)$ increase with decreasing $p$, again in agreement with this scenario. As noted earlier~\cite{Bakr_PRB2013}, $T_{CDW}(p)$ always remains below $T^*(p)$ and goes through a maximum at $p \sim 0.12$, whereas $T^*(p)$ increases monotonically with decreasing $p$. This confirms that the CDW correlations are not the root cause of the pseudogap phenomenon, and that at least for low to moderate doping the pseudogap cannot be thought of as a CDW gap. \cite{Alloul_arxiv} Rather, the CDW must be regarded as an instability inside the pseudogap regime. We also point out that the temperature dependence of the CDW correlations reported here does not track the one of the polar Kerr effect~\cite{Xia_PRL2008,Kapitulnik_NJP2009} and the $q=0$ magnetic order detected by neutron diffraction \cite{Fauqué_PRL2006, Li_Nature2008, Baledent_PRB2011}, leaving open at this stage the relationship between the CDW fluctuations and these effects~\cite{Hosur_PRB2013}.

On the other hand, it is interesting to note the close similarity of $T_{CDW}(p)$ and the onset of intra-bilayer superconducting fluctuations in YBCO$_{6+x}$ inferred from the $c$-axis optical conductivity, which also exhibits a dome-like shape in the $p-T$ diagram, with a shallow peak for $p \sim 0.1$ and $T \sim 180$ K. \cite{Dubroka_PRL2011}. The combined onset of superconducting and CDW fluctuations and the wide fluctuation regime suggest a composite order parameter subject to strong phase fluctuations~\cite{Hayward}, consistent with recent proposals of a fluctuating pair density wave (PDW) state~\cite{Berg_PRL2007, Berg_NJP2009, Nie_PNAS2014, Lee_PRX2014, Corboz_PRL2014}.
While $d$-wave superconductivity preempts PDW long-range order for $H = 0$, recent magnetometric experiments suggest that it may be the leading instability in high magnetic fields. \cite{Yu_arXiv}

\subsection{In-plane anisotropy and relationship to stripes}

The in-plane anisotropy provides further information about the microscopic structure of the CDW state. We have shown that this quantity evolves systematically as a function of doping (Fig.~\ref{fig:REXS_a_vs_b}). For $p \sim 0.12$, where the intensity, correlation length, and onset temperature of the CDW correlations are maximal, the RXS peaks along the $h$- and $k$-directions have approximately the same amplitude. This is most easily understood in terms of a biaxial ``checkerboard'' modulation, although an accidental, approximately equal mixture of uniaxial domains cannot be ruled out based on the data presented here. Away from this doping level, the in-plane anisotropy increases. For $p \rightarrow p_{c1}$, the peak along $h$ disappears below the detection limit, consistent with a uniaxial modulation. For $p \rightarrow p_{c2}$, on the other hand, the RXS peaks along $k$ become more intense.

One might be tempted to associate the in-plane anisotropy of the CDW and its $p$-evolution with changes in the electronic structure induced by the commensurate ordering of oxygen dopant atoms in the chain layer. In this case, however, one would expect that the degradation of ortho-II order for $x < 0.5$ restores the
isotropy of the CDW peaks, which is clearly not the case. Indeed, the anisotropy persists down to the lowest doping level where the CDW is observed. The apparent $p$-induced sign reversal of the intensity anisotropy is also difficult to attribute to oxygen order in the chains. While an influence of the ortho-I structure on the modest anisotropy at high $p$ cannot be ruled out, the large anisotropy for $p \rightarrow p_{c1}$ appears to be a consequence of an intrinsic tendency towards uniaxial CDW order in the CuO$_2$ planes. A related trend is observed in the spin fluctuation spectrum as $p_{c1}$ is approached from below.~\cite{Haug_NJP2010} Note, however, that the propagation vector of the incommensurate spin fluctuations is along $h$, that is, perpendicular to the soft charge fluctuations for $p > p_{c1}$.

Unless the isotropy of the CDW for $p \sim 0.12$ is purely accidental, it thus appears that the fluctuations in the center of the CDW ``dome'' (Fig.~\ref{fig:Tonset}) are biaxial, whereas those for both lower and higher doping are increasingly uniaxial. Qualitatively, this situation resembles the recently investigated phase diagram of helical magnets in magnetic fields, which include both single-$q$ (spiral) and multiple-$q$ (skyrmion lattice) states.~\cite{Adams_PRL2012,Seki_PRB2012} In the cuprates, the distinction between single- and double-$q$ CDW structures is possibly blurred by disorder. We note, however, that sound velocity measurements performed on a ortho-II ordered YBCO$_{6.55}$ single crystal under large magnetic field indicate a biaxial modulation~\cite{LeBoeuf_NaturePhysics2013}. Moreover, according to model calculations only a biaxial modulation can induce a Fermi surface reconstruction compatible with the experimentally observed period of the quantum oscillations~\cite{Sebastian_PRL2012}. Surprisingly, quantum oscillations have been reported at high magnetic fields even in the regime $0.086 \leq p \leq 0.1$ where our low-field data indicate uniaxial CDW order (Fig.~\ref{fig:REXS_a_vs_b}), with no qualitative differences to the regime with more isotropic CDW correlations at higher $p$. At this stage, it cannot be excluded that the long-range ordered CDW state for high $H$ differs from the short-range ordered state at lower $H$. Julien \textit{et al.} have suggested that a transition between two CDW states occurs at $H \sim 20$ T.~\cite{wu_arxiv} At the present time, this field is difficult to access with scattering probes.

We now discuss in more detail the relationship between the charge modulations we have described in the 123 system to those in the 214 system, which have been widely discussed in terms of uniaxial (``stripe'') modulations.
On the one hand, the charge correlations in 123 and 214 exhibit quantitative and qualitative differences. The most striking difference is the opposite doping dependence of $\delta_{CDW}$ in the two families (Fig.~\ref{fig:REXS_DopingDep}). In 214 compounds~\cite{Hucker_PRB2011,Fink_PRB2009,Fink_PRB2011}, it is well established that the wave vector of the charge modulation increases with doping (Fig. ~\ref{fig:QCDW}), in lockstep with the $p$-dependence of the spin correlations with incommensurability $\delta_{SDW} = \delta_{CDW}/2$, \cite{Fujita_JPSJ2012} and then saturates near the commensurate value $\delta_{CDW} = 1/4$. In the 123 system, the wave vector characterizing the quasi-static, nearly antiferromagnetic spin correlations close to the Mott-insulating state ($p \lesssim 0.08$) also increases with increasing doping. \cite{Haug_NJP2010} Since a similar trend has recently been established for the nearly critical spin fluctuations in Bi$_2$Sr$_2$CuO$_{6+\delta}$ at low $p$, \cite{Enoki_PRL2013} this behavior can be regarded as universal for the deeply underdoped regime of the cuprate phase diagram.

Whereas coupled, nearly critical spin and charge fluctuations persist in 214 over a wide range of $p$, continuing the trend that can be traced back to the Mott-insulator, the 123 system goes through a sharply defined critical point that separates regimes with ungapped ($p < p_{c1}$) \cite{Haug_NJP2010} and gapped ($p > p_{c1}$) \cite{Fong_PRB2000} spin fluctuations. For $p > p_{c1}$, nearly critical charge fluctuations appear at a wavevector that is tied to the evolution of the Fermi surface \cite{Comin_Science2014, Tabis}. This behavior is also observed for the Bi- and Hg-based cuprates~\cite{Comin_Science2014,daSilvaNeto_Science2014,Hashimoto_PRB2014,Tabis}, and can hence be regarded as generic for the moderately doped cuprates.

On the other hand, we have already noted that the momentum-integrated amplitudes of the CDW peaks in both materials are comparable \cite{Thampy_PRB2013}. Comparison between our current measurements and prior RXS work on La$_{2-x}$Ba$_x$CuO$_4$ (Ref.~\onlinecite{Hucker_PRB2011}) and La$_{1.8-x}$Eu$_{0.2}$Ba$_x$CuO$_4$ (Ref.~\onlinecite{Fink_PRB2011}) now shows that the CDW stability ranges (Fig.~\ref{fig:REXS_DopingDep}) are also remarkably similar in the 123 and 214 systems, as are the pronounced maxima of the amplitude, correlation length, and onset temperature of the charge-ordering reflections for $p \sim 0.12$. In both cases, the superconducting $T_c$ is reduced in this range of doping levels, at least with respect to the quadratic $T_c$-versus-$p$ relation that can be extrapolated from lower and higher $p$~\cite{Tallon_PRB1995}. Furthermore, we note that there is a striking analogy between the anomalies in the phonon dispersions~\cite{Reznik_PRB2008,Reznik_Nature2006, Raichle_PRL2011} and transport properties~\cite{Sebastian_PNAS2010, Laliberte_NatCom2011} associated with the charge modulations in both families of compounds. In the 214 system, the onset temperature of charge ordering is generally higher than the one for spin ordering~\cite{Fink_PRB2011,Hucker_PRB2011}.
Charge ordering appears therefore as the leading instability competing with superconductivity in all underdoped cuprates around $p \sim 1/8$. The presence (or absence) of spin ordering at lower temperatures has been described in the framework of a Landau theory of coupled charge and spin density wave order parameter.~\cite{Zachar_PRB1998}.
The apparent non-generic behavior in the 214 system (including the lock-in of the CDW wavevector to the commensurate value of 1/4 for $p \sim 1/8$) may therefore be a consequence of microscopic details, such as the tilt distortions of the CuO$_6$ octahedra in the 214 lattice structure (and not observed in other cuprate families), 
or disorder due to the randomly placed Sr/Ba donors. Those may also contribute to the stabilization of incommensurate magnetism in the 214 systems, in analogy with the effect of nonmagnetic impurities in moderately doped 123, that were shown to close the spin gap and to induce incommensurate magnetic order at the expense of CDW correlations~\cite{Alloul_RMP2009,Suchaneck_PRL2010,Blanco_PRL2013}.

\section{Conclusion}
\label{sec:conclusion}

We have detected incommensurate charge density wave fluctuations in YBCO$_{6+x}$ for hole doping levels $p_{c1} \leq p \leq p_{c2}$ with $p_{c1} \sim 0.08$ and $p_{c2} \sim 0.18$. The onset temperature of the CDW correlations forms a ``dome'' ranging from $p_{c1}$ to $p_{c2}$ in the $p-T$ phase diagram, with a peak of $T_{CDW} \sim 160$ K for $p \sim 0.12$. The peak temperature coincides with the onset of superconducting fluctuations detected by infrared spectroscopy, \cite{Dubroka_PRL2011} and with the mean-field transition temperature for $d$-wave superconductivity calculated based on the experimentally observed spin fluctuation spectrum \cite{Dahm_NaturePhysics2009,letacon_NatPhys2011}. These findings suggests strong, combined fluctuations of the $d$-wave superconducting and CDW order parameters, and they are consistent with proposals of a proximate ground state with a composite order parameter (such as the pair density wave) that competes with the uniform $d$-wave pairing state and generates the plateau in the $T_c$-versus-$p$ relation ~\cite{Berg_PRL2007, Berg_NJP2009, Nie_PNAS2014, Lee_PRX2014, Corboz_PRL2014}. Further work is required to establish whether such a state becomes thermodynamically stable in high magnetic fields.

The temperature and magnetic field dependence of the RXS intensity for fixed $p$, combined with recent high-field transport experiments, suggest proximate CDW quantum critical points for $p = p_{c1}$ and $p_{c2}$. For $p \lesssim p_{c1}$, soft incommensurate spin fluctuations set in, and $T_c$ is further reduced, consistent with spin fluctuation mediated pairing models.  \cite{Dahm_NaturePhysics2009,letacon_NatPhys2011} The presence of similar, ungapped spin fluctuations over a wide doping range may be responsible for the lower maximal $T_c$ in the 214 system. Remarkably, $p_{c2}$ is close to the doping level optimal for superconductivity, and according to an influential study, \cite{Tallon_PhysicaC2001} with the end point of the pseudogap regime inside the superconducting dome. In high fields, superconductivity is suppressed most strongly for $p \sim 0.12$, and $p_{c1}$ and $p_{c2}$ become centers of separate superconducting domes~\cite{Grissonnanche2014}. The possible role of quantum-critical CDW fluctuations for the mechanism of high-$T_c$ superconductivity suggested by these observations is an important subject of future experimental and theoretical research.

\noindent \textit{Note added}. H\"{u}cker {\it et al.} have recently reported similar results on the doping dependence of the CDW in YBCO using hard x-rays diffraction.~\cite{Huecker_2014}

\section*{Acknowledgements}

We acknowledge M.-H. Julien, B. Ramshaw and S. Sebastian for fruitful discussions and sharing with us unpublished data, S. A. Kivelson and P. A. Lee for critical and insightful reading of the manuscript and C.T. Lin for sample preparation.

\end{document}